\documentclass[noshowpacs,amsmath,
twocolumn,
superscriptaddress,
8pt,
aps,prb]{revtex4-1}
\usepackage{setspace}
\usepackage{amsmath}
\usepackage{graphicx}
\usepackage[nearskip,margin = 0pt]{subfig}

\usepackage{verbatim}
\usepackage{amsfonts}
\usepackage{amssymb}
\usepackage{epstopdf} 
\usepackage{xcolor}
\usepackage{verbatim}
\usepackage{upgreek}
\usepackage{ragged2e}
\usepackage{url}
\usepackage[hidelinks]{hyperref}
\DeclareGraphicsExtensions{.pdf,.eps,.png,.jpg,.mps}

\begin{document}
\title{Photonic chip-based optical frequency division with PZT-integrated soliton microcombs}



\author{Ruxuan Liu$^{1}$, Mark W. Harrington$^{2}$, Shuman Sun$^{1}$, Fatemehsadat Tabatabaei$^{1}$, Samin Hanifi$^{1}$, Meiting Song$^{2}$, Kaikai Liu$^{2}$, Jiawei Wang$^{2}$, Haoran Chen$^{1}$ , Zijiao Yang$^{1,3}$, Beichen Wang$^{1}$, Fateme Majdi$^{1}$, Paul A. Morton$^{4}$, Karl D. Nelson$^{5}$, Steve M. Bowers$^{1}$, Andreas Beling$^{1}$, Daniel J. Blumenthal$^{2,\dagger}$, and Xu Yi$^{1,3,\dagger}$\\
\vspace{3pt}
$^1$Department of Electrical and Computer Engineering, University of Virginia, Charlottesville, Virginia 22904, USA.\\
$^2$Department of Electrical and Computer Engineering, University of California Santa Barbara, Santa Barbara, California 93016, USA.\\
$^3$Department of Physics, University of Virginia, Charlottesville, Virginia 22904, USA.\\
$^4$Infleqtion, Louisville, CO, 80027, USA.\\
$^5$Honeywell International, Plymouth, Minnesota, USA.\\
$^{\dagger}$Corresponding authors: yi@virginia.edu, danb@ucsb.edu}

\begin{abstract}
Optical frequency division (OFD) produces low-noise microwave and millimeter-wave signals by transferring the exceptional stability of optical references to electronic frequency domains. Recent developments in integrated optical references and soliton microcombs have paved the way for miniaturizing OFD oscillators to chip scale. Critical to this realization is a rapid tunable frequency comb that is stabilized to the optical references, thereby coherently linking optical and electronic frequencies. In this work, we advance the on-chip OFD technology using an integrated high-speed PZT stress-optic actuator on the SiN soliton microcomb resonator. The integrated PZT actuator tunes the resonance frequency of the soliton-generating microresonator with a bandwidth exceeding 10s MHz and independently adjusts the soliton repetition rate without perturbing the frequency comb offset.
Optical frequency division and low-noise mmWave generation are demonstrated by feedback control of the soliton repetition rate through the integrated PZT-actuator, and the soliton microcomb is stabilized to a pair of reference lasers that are locked to an integrated 4-meter SiN coil reference cavity. Our approach provides a fast, versatile and integrated control mechanism for OFD oscillators and their applications in advanced communications, sensing, and precise timing.


\end{abstract}
\date{\today}

\maketitle

\noindent {\bf Introduction}



Optical frequency division (OFD) is able to transfer the exceptional fractional stability from stable optical references to electronic frequencies through the use of optical frequency combs (OFCs)\cite{fortier2011generation}.
It has set the phase noise record in microwave and mmWave regimes \cite{xie2017photonic,nakamura2020coherent,tetsumoto2021optically,li2023small}, with the potential to revolutionize many applications, including communications, sensing, and precise timing \cite{koenig2013wireless,ghelfi2014fully,long2015microwave}. 
Very recently, OFD technology has been transitioning to chip scale thanks to developments in integrated optical references \cite{lee2013spiral,jin2021hertz,li2021reaching,liu202236,guo2022chip} and soliton microresonator-based frequency combs (microcombs) \cite{herr2014temporal,yi2015soliton,brasch2016photonic,kippenberg2018dissipative,gaeta2019photonic}, and is rapidly advancing the spectral purity of integrated microwave and millimeter-wave oscillators\cite{sun2024integrated,kudelin2024photonic,zhao2024all,he2024chip,sun2025microcavity,ji2025dispersive,jin2025microresonator}. 

The implementation of OFD relies upon the feedback control of an optical frequency comb, which stabilizes the comb to the optical reference and provides a coherent link between optical and electronic frequencies. A direct, rapid tuning mechanism for the frequency comb cavity is thus critical to OFD oscillators. The miniaturization of OFD oscillators offers a unique opportunity to enable a much faster control on the frequency comb than that in the bulk OFD systems, as the bandwidth of electronics and mechanics typically increases with the decrease of device dimension. Indeed, integrated stress-optics actuators, such as PZT actuators, have reached a modulation bandwidth of 10s MHz \cite{hosseini2015stress,jin2018piezoelectrically} and are readily integrated with high-Q optical microcavities \cite{liu2020monolithic,lihachev2022low,wang2022silicon}. Electro-optic modulators provide even higher tuning speed\cite{wang2018integrated,he2023high}.
However, their integration into OFD oscillators has not been explored in previous on-chip OFD demonstrations, where the microcomb was tuned either by the frequency of its pump laser\cite{tetsumoto2021optically,sun2024integrated,kudelin2024photonic,ji2025dispersive,jin2025microresonator}, or passively using laser injection locking \cite{taheri2017optical,moille2023kerr,wildi2023sideband,zhao2024all,sun2025microcavity}. 

\medskip

\begin{figure*}[!bht]
\captionsetup{singlelinecheck=off, justification = RaggedRight}
\includegraphics[width=17cm]{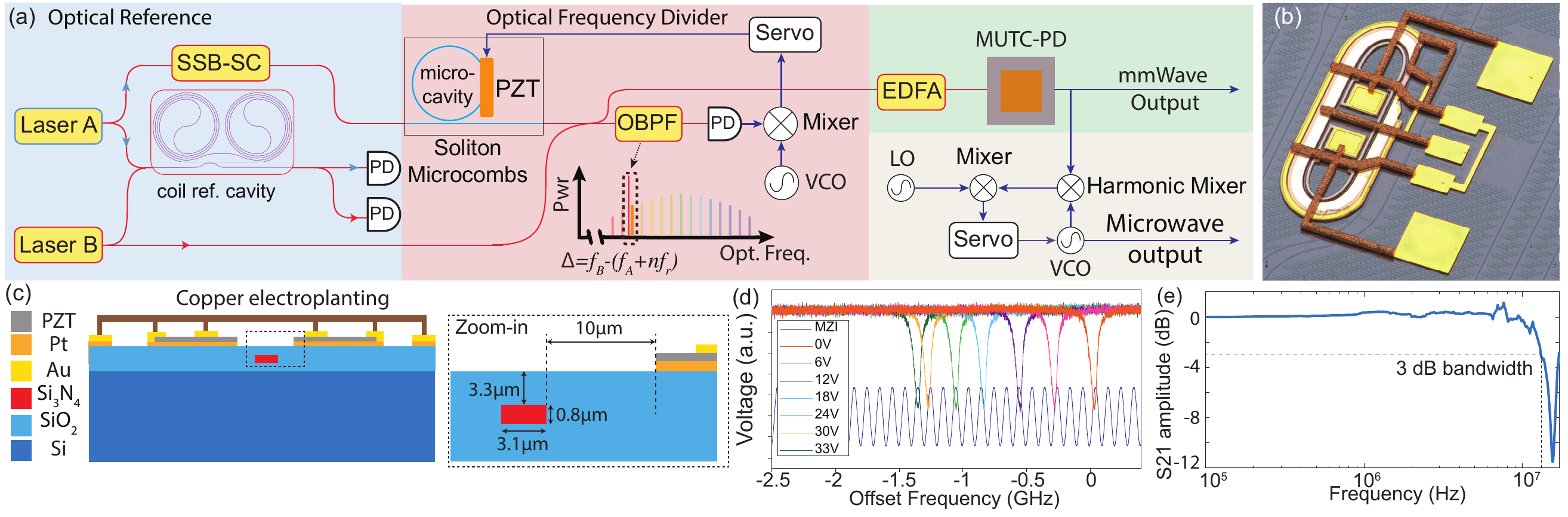}
\caption{{Conceptual illustration of OFD with PZT actuators and characterization of PZT-integrated microresonator.}
    \textbf{(a)}  Simplified illustration for the OFD experimental setup. A pair of lasers at 1550 nm (laser A) and 1600 nm (laser B) are stabilized to an integrated 4-meter SiN coil reference cavity using the PDH locking technique. A fast-tuning sideband of laser A, created by modulating a single-sideband suppressed carrier modulator (SSB-SC), is used to generate the soliton microcomb in the PZT-integrated SiN microcavity. To implement the OFD, the $N$-th comb line is phase locked to laser B by feedback control of the soliton repetition rate using the integrated PZT actuator. The soliton microcomb is then amplified and photodetected on a high-speed MUTC-PD to generate low noise mmWave. Erbium-doped fibre amplifiers (EDFAs), voltage controlled oscillators (VCOs), optical bandpass filter (OBPF), and photodiodes (PDs) are used in the experiment. \textbf{(b)} 3D illustration of a typical PZT-integrated SiN microcomb resonator. \textbf{(c)} Illustrative cross section of the SiN waveguide and PZT actuator layers. Not drawn to scale. Zoomed-in panel showing on the right. \textbf{(d)} Microresonator resonance frequency tuning versus bias voltage on PZT. An MZI with 100 MHz period is used for frequency calibration. \textbf{(e)} Frequency response (S21) to characterize the bandwidth of the PZT actuator.}
\label{fig:1}
\end{figure*}

In this work, we demonstrate optical frequency division (OFD) and low-noise mmWave generation by using a PZT-integrated soliton microcomb. The PZT-based stress-optics actuator is directly integrated on the SiN soliton microcomb cavity, and it can tune the microcavity resonance frequency and thus the repetition rate of the soliton microcomb. The actuator has a 3-dB bandwidth of 10s MHz, and the speed of the OFD feedback control is limited by the electronics of the PID loop. The PZT-integrated soliton microcomb is stabilized to a pair of reference lasers that are PDH locked to an integrated SiN 4-meter coil reference cavity\cite{liu202236}. 109.5 GHz mmWave signal is generated by photodetecting the soliton microcombs with a modified uni-traveling carrier photodiode (MUTC-PD)\cite{wang2021towards}. The phase noise at 10 kHz offset frequency reaches -114 dBc/Hz, which is equivalent to -135 dBc/Hz when the carrier frequency is scaled to 10 GHz. The phase noise performance aligns well with our previous record-setting OFD demonstrations \cite{sun2024integrated,sun2025microcavity} at 100s GHz carrier frequency, and is primarily limited by the optical reference and mmWave phase noise measurement method.

\medskip





\noindent {\bf Results.}

A PZT-integrated $Si_3N_4$ racetrack-shaped microresonator is used for soliton microcomb generation. The resonator has a cross-section of 3.1 $\mu$m width $\times$ 0.8 $\mu$m height, a free spectral range (FSR) of 109.5 GHz and an intrinsic (loaded) quality factor of 3.85 $\times 10^{6}$ (2.87 $\times 10^{6}$). The SiN resonator has a top oxide cladding layer of 3.3 $\mu$m. The piezo-electric (PZT, lead zirconate titanate) actuator is then fabricated on top of the oxide cladding. Details of the PZT actuator fabrication have been described elsewhere\cite{wang2022silicon}. A side-view of the PZT actuator structure is shown in Figure.\ref{fig:1}c. For SiN waveguide of 800 nm thickness, the optical mode is well confined in the SiN core, and the PZT actuator will not impact the Q-factor of the microresonator. 

The modulation property of the PZT-integrated microresonator is characterized and presented in Fig.\ref{fig:1}d,e. The PZT actuator is tuned by applying an electrical field using a DC probe, and the waveguide refractive index is changed by the strain induced through the piezoelectric effect. The frequency shift of the microresonator in response to the applied DC voltage on the PZT probe is shown in Fig. \ref{fig:1}d. A tuning coefficient of 43.7 MHz/Volt is observed, with the accuracy limited by the relative frequency drift of the resonator and the calibrating Mach–Zehnder interferometer (MZI) due to environmental temperature fluctuation. The small-signal electrical-to-optical modulation response S21 is measured with a vector network analyzer (VNA). The small signal is sent to the PZT actuator, and the transmission of the microresonator is detected on a photodiode. To obtain the information of the resonance frequency, a frequency to amplitude conversion is required, which is achieved by detuning the probe laser frequency to the full-width-half-maximum (FWHM) point of the microcavity resonance. The S21 magnitude is shown in Fig. \ref{fig:1}e. A 3-dB bandwidth is measured to be around 13 MHz. The small-signal measurement is sensitive to the relative detuning of the probe laser and the resonator frequency. It should be noted that a resonance dip is observed around 20 MHz frequency, which likely comes from the electronic circuits instead of the PZT itself.  Nevertheless, this measured bandwidth far exceeds the servo bandwidth achieved in OFD\cite{sun2024integrated,kudelin2024photonic,jin2025microresonator}, and will not be a limiting factor of the OFD bandwidth.


\begin{figure*}[!bht]
\captionsetup{singlelinecheck=off, justification = RaggedRight}
\includegraphics[width=17cm]{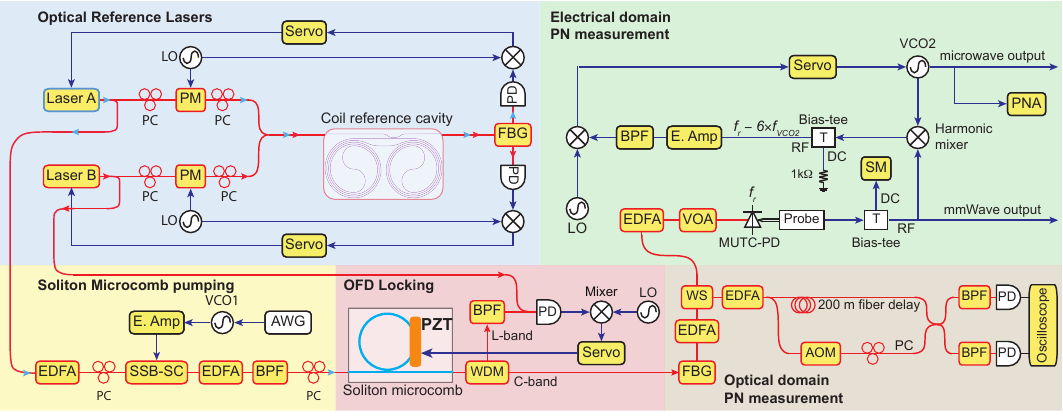}
\caption{{\bf Full experimental setup.}
    \textbf{(a)} The PZT-OFD oscillator outputs a stable mmWave signal at 110 GHz. The mmWave phase noise is characterized by a dual-tone delayed self-heterodyne interferometry method (optical domain) and a mmWave to microwave frequency division (mmFD) method (electrical domain). In the electrical method, the mmFD down-converts the carrier frequency from 110 GHz to 18 GHz for direct measurement on a phase noise analyzer (PNA). Polarization controllers (PCs), phase modulators (PMs), bandpass filters (BPFs), fibre-Bragg grating (FBG) filters, wavelength division multiplexer (WDM), line-by-line waveshaper (WS), electrical amplifier (E. Amps), source meter (SM), and acoustic-optics modulator (AOM) are used in the experiment.}
\label{fig:2}
\end{figure*}

The PZT-actuated soliton microcomb is then applied to demonstrate optical frequency division and low-noise mmWave generation. In our OFD demonstration, the oscillator derives its stability from two optical references near 1550 nm ($f_A = 193.5$ THz) and 1600 nm ($f_B=187.4$ THz), which are both stabilized to a common integrated 4-meter SiN coil reference cavity \cite{liu202236} through the Pound-Drever-Hall (PDH) locking technique. A modulation sideband of reference laser A directly drives the soliton microcomb\cite{stone2018thermal} and serves as 0-th comb line, with the comb line frequency at:  $f_0 = f_A + f_\text{VCO1}$. The reference lasers and soliton generation have been described elsewhere\cite{sun2024integrated,sun2025microcavity}. To implement OFD, the second reference laser at 1600 nm is photomixed with the comb line at $N=-54$, and their beatnote is used to phase lock the $N$-th comb line to the reference laser by feedback control of soliton repetition rate through tuning the PZT actuator. No optical amplifier is used to amplify the power of the comb line for OFD locking.
When phase locked, the frequency of the $N$ -th comb line will be $f_N = f_B + f_\text{VCO2}$, which also equals $f_N = f_0 + N\times f_r = f_A + f_\text{VCO1} + N\times f_r$, where $f_r$ is the repetition rate of the soliton. An optical frequency division is thus achieved, as $f_r = (f_B-f_A)/N + (f_\text{VCO2}-f_\text{VCO1})/N$. In our system, the phase noise of the two VCOs is much lower than that of the two optical reference lasers, and thus the phase noise of soliton repetition rate is dominated by the divided optical references: $S_r = (S_A + S_B)/N^2$, where $S_r$, $S_A$ and $S_B$ are the phase noise of soliton repetition rate, reference laser A and reference laser B, respectively. This configuration achieves a frequency division ratio of 54 (from 6 THz to 110 GHz), offering a phase noise reduction of $54^2 \approx 35$ dB.

\begin{figure*}[!bht]
\captionsetup{singlelinecheck=off, justification = RaggedRight}
\includegraphics[width=17cm]{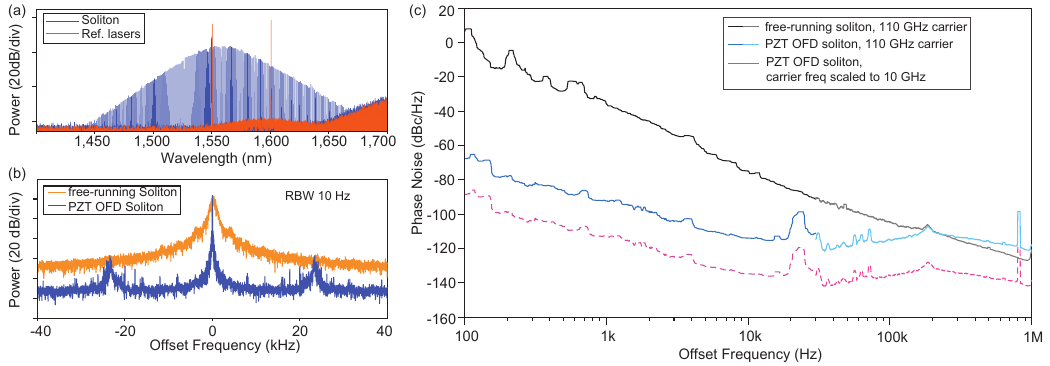}
\caption{{\bf Results of optical frequency division using PZT-integrated soliton microcomb. } \textbf{(a)} Optical spectra of soliton microcomb (blue) and a pair of reference lasers (orange). \textbf{(b)} The electrical spectra of the mmWave output after mmWave to microwave frequency division for free running soliton (orange) and PZT-integrated OFD soliton (blue). \textbf{(c)} Phase noise characterization of free-running soliton (black) and integrated-PZT OFD soliton (blue). The pink trace shows the phase noise of PZT OFD soliton when the carrier frequency is scaled down to 10 GHz.}
\label{fig:3}
\end{figure*}
 
The phase-stabilized PZT-controlled solitons are then amplified and sent to a high-speed MUTC photodiode to generate a low-noise mmWave signal at 110 GHz carrier frequency \cite{wang2021towards}. To measure the phase noise of 110 GHz, two methods are utilized. For offset frequencies below 30 kHz, a mmWave to microwave frequency division (mmFD)\cite{sun2024integrated} is applied to down-convert the carrier frequency from 110 GHz to 18 GHz, whose phase noise and electrical spectra (Fig. \ref{fig:3}b) can be directly measured on phase noise analyzer \cite{sun2024integrated}. The measured phase noise is then scaled back to 110 GHz to give an upper bound limit of the mmWave phase noise. The mmFD method has a locking bandwidth of roughly 282 kHz. Beyond 30 kHz offset frequency, the phase noise is measured optically using a dual-tone delayed self-heterodyne interferometry method\cite{kwon2017reference,sun2024integrated}. The combined phase noise is shown in Fig.\ref{fig:3}c, with carrier frequency scaled to 110 GHz (native) and 10 GHz.

The phase noise of the generated mmWave is very low for an integrated oscillator. At 10 kHz offset, the oscillator has a phase noise of -114 dBc/Hz, which is equivalent to -135 dBc/Hz when the carrier frequency is scaled down to 10 GHz. This result is 2 dB better than our non-PZT integrated OFD oscillator reported previously \cite{sun2024integrated}. The phase noise of both oscillators is primarily limited by the reference lasers. Also, the mmFD phase noise measurement at 100s GHz contributes excess phase noise. Surprisingly, the PZT OFD servo bandwidth is only 200 kHz, which is significantly lower than the bandwidth of PZT itself. We believe that this is limited by the servo loop itself, and could be dramatically improved in the future when electronics and feedback loop are specifically designed for PZT-enabled OFD.

\medskip
\noindent {\bf Summary.}
In summary, we have demonstrated optical frequency division and low noise mmWave generation using PZT-integrated soliton microcombs. The integrated PZT actuator contributes a new rapid tuning mechanism in OFD oscillators and simplifies its implementation. In our system, only two reference lasers are used to both generate the solitons and implement OFD. No additional laser is required to pump the soliton and tune its repetition rate, which is much improved from the initial integrated OFD demonstration\cite{sun2024integrated,kudelin2024photonic}. Compared with the passive locking OFD method, e.g., Kerr OFD \cite{moille2023kerr,zhao2024all,sun2025microcavity}, the PZT OFD does not require fine frequency alignment of the reference lasers and microcombs. Finally, in our demonstration, in addition to optical frequency division, the entire soliton microcomb is fully phase stabilized to the reference lasers, as two comb lines, the 0-th line and $N$-th line, are stabilized to the reference lasers. This eliminates any free-running parameter in the frequency of the microcomb, and thus achieves full stabilization to a reference cavity.







\begin{footnotesize}

\end{footnotesize}

\medskip

\noindent\textbf{Acknowledgement}

\noindent The authors acknowledge Ryan Rudy from U.S. Army Research Laboratory for the PZT actuator fabrication, M. Woodson and S. Estrella from Freedom Photonics for the MUTC PD fabrication, Ligentec for SiN microresonator fabrication, and gratefully acknowledge DARPA GRYPHON (HR0011-22-2-0008), DARPA NaPSAC (N660012424000) and National Science Foundation (2023775). The views and conclusions contained in this document are those of the authors and should not be interpreted as representing official policies of DARPA or the U.S. Government.

\medskip

\noindent \textbf{Author Contributions}\\ 
X.Y. and R.L designed the experiments. R.L, F.T. and S.H. performed the oscillator measurements. R.L and M.W.H characterized the PZT. J.S.M. and A.B. designed and fabricated the integrated photodiodes. S.S., Z.Y., B.W. designed the integrated photonic circuits. R.L and X.Y. analyzed the experimental results. X.Y., D.J.B., R.R., A.B., S.M.B., P.A.M., and K.D.N. supervised and led the scientific collaboration. All authors participated in preparing the manuscript.\\


\medskip

{\noindent \bf Competing interests}
The authors declare no competing interests.

\medskip

{\noindent \bf Data availability.} The data that support the plots within this paper and other findings of this study are available from the corresponding author upon reasonable request. Accession to all relevant data will be available online before publication.


\medskip

{\noindent \bf Code availability.} The codes that support the findings of this study are available from the corresponding authors upon reasonable request.

\bibliographystyle{naturemag}
\bibliography{ref}

\end{document}